\documentclass[prb,twocolumn,aps,preprintnumbers,amsmath,amssymb]{revtex4}
\usepackage{graphicx}

\begin{document}
\preprint{}
\draft

\title{Robust photon-spin entangling gate using a quantum-dot spin in a microcavity}

\author{C.Y.~Hu$^{1}$}\email{chengyong.hu@bristol.ac.uk}
\author{W.J.~Munro$^{2,3}$}
\author{J.L.~O'Brien$^1$}
\author{J.G.~Rarity$^1$}
\affiliation{$^1$Department of Electrical and Electronic Engineering, University of Bristol, University Walk, Bristol BS8 1TR, United Kingdom}
\affiliation{$^2$Hewlett-Packard Laboratories, Filton Road, Stoke Gifford, Bristol BS34 8QZ, United Kingdom}
\affiliation{$^3$National Institute of Informatics, 2-1-2 Hitotsubashi, Chiyoda-ku, Tokyo 101-8430, Japan}

\begin{abstract}

Semiconductor quantum dots (known as artificial atoms) hold great promise for solid-state quantum networks and quantum computers.
To realize a quantum network, it is crucial to achieve light-matter entanglement and coherent quantum-state transfer between light and matter.
Here we present a robust photon-spin entangling gate with high fidelity and high efficiency (up to 50 percent) using a charged quantum dot
in a double-sided microcavity. This gate is based on giant circular birefringence induced by a single electron spin, and functions as an
optical circular polariser which allows only one circularly-polarized component of light to be transmitted depending on the
electron spin states. We show this gate can be used for single-shot quantum non-demolition measurement of a single electron spin,
and can work as an entanglement filter to make a photon-spin entangler, spin entangler and photon entangler as well as a photon-spin
quantum interface. This work allows us to make all building blocks for solid-state quantum networks with single photons and
quantum-dot spins.

\end{abstract}

\date{\today}


\maketitle

\section{Introduction}

A quantum network \cite{cirac97}  utilizes matter quantum bits (qubits) to store and process quantum information at local nodes,
and light qubits (photons) for long-distance quantum state transmission between different nodes. Quantum networks can be used for distributed
quantum computing or for large scale and long distance quantum communications between spatially remote parties. There are several physical
systems based on cavity quantum electrodynamics (cavity-QED), which could be used for quantum networks with  high success probability for
quantum-state transfer or processing. One is the atom-cavity system in which single photon sources \cite{kuhn02, mckeever04, wilk07a},
light-atom entanglement \cite{sherson06, boozer07} and a single-photon single-atom quantum interface \cite{wilk07} have been recently
demonstrated. But it is far from a trivial task to scale up and to trap the atoms. Another one is the superconducting qubit-cavity system
which attracts great interest in recent years.  Single photon generation in the microwave frequency region \cite{houck07}  and a quantum
bus allowing distant qubits to interact at will \cite{sillanpaa07, majer07} have been implemented recently in this system.

The third one is the semiconductor quantum dot (QD)-cavity system \cite{loss98, imamoglu99, calarco03, yao05, clark07}. Firstly,
triggered single-photon sources or polarization-entangled photon pair sources based on semiconductor QDs have been demonstrated with
high quantum efficiency, high photon indistinguishability, and low multi-photon emission
probability \cite{moreau01, yuan02, santori02, stevenson06, akopian06}.
These  deterministic photon sources  are key ingredients for secure quantum networks. Secondly, semiconductor QD spins are promising candidates to
construct qubits for storing and processing quantum states \cite{awschalom02, atature06} due to the long electron spin coherence
time ($T_2\sim \mu$s) \cite{petta05, greilich06} and  spin relaxation time ($T_1\sim$ms) \cite{kroutvar04}.
Moreover, self-assembled QDs can be embedded in various high-finesse optical microcavities or nanocavities, so cavity-QED can be
exploited to engineer QD emissions or related optical transitions as demanded \cite{reithmaier04, yoshie04, peter05, hennessy07,
press07, reitzenstein07}. The most attractive feature is its compatibility with standard semiconductor processing techniques.
Therefore, the QD-cavity system holds great promise for compact and scalable solid-state quantum networks and quantum computers.
However, the photon-spin entanglement and quantum state transfer between photon and QD spin  have not yet been demonstrated \cite{yao05, flindt07}.

Here we propose a robust photon-spin entangling gate using a charged QD in a double-sided microcavity, and show this gate can be used as  photon-spin
entangler, spin entangler, photon entangler as well as reversible and coherent quantum-state transfer between single photons and QD spins.
This gate is based on giant circular birefringence induced by a single electron spin, and is ideal  for an optical quantum
non-demolition (QND) measurement of a single electron spin in a double-sided microcavity. This gate is robust and flexible compared to our previous
gate using a charged QD in a single-sided microcavity \cite{hu08a, hu08b}.

The paper is organized as follows: In Sec.II, the photon-spin
entangling gate is introduced. In Sec. III we show this gate can be used for single-shot QND measurements of a single QD spin.
After that, we show a spin entangler in Sec. IV, a photon entangler in Sec. V and a photon-spin quantum interface in Sec. VI
by applying this photon-spin entangling gate. Finally, we present our conclusions and outlook in Sec. VII.

\section{Photon-spin entangling gate}

We consider a singly charged QD, e.g., a self-assembled In(Ga)As QD or a GaAs interfacial QD, or even a semiconductor nanocrystal
inside an optical cavity, such as a micropillar or microdisk microcavity, or a photonic crystal nanocavity. Fig. 1a shows a  micropillar
microcavity where the two GaAs/Al(Ga)As distributed Bragg reflectors (DBR)  and the transverse index guiding
provide the three-dimensional confinement of light. The two DBRs are made symmetric in order to achieve high resonant transmission of
light. Both DBRs are partially reflective allowing light into and out of the cavity (i.e., a double-sided cavity). The circular cross section of
the micropillar supports the circularly polarized light. The QD is located at the antinodes of the cavity field to achieve optimized
light-matter coupling.

\begin{figure}[ht]
\centering
\includegraphics* [bb= 104 202 500 625, clip, width=6cm,height=6cm]{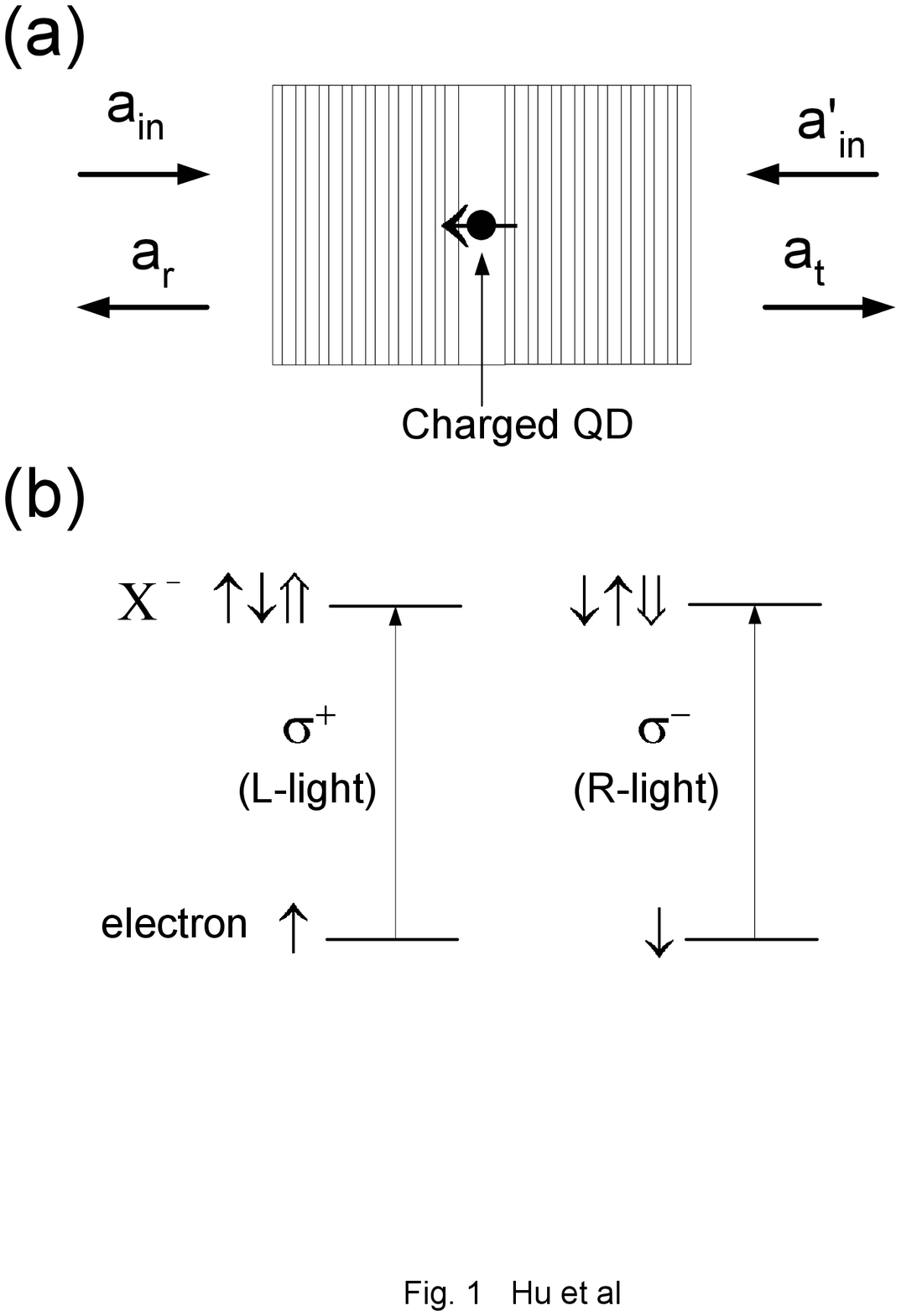}
\caption{(a) A charged QD inside a micropillar microcavity with circular cross section.
(b) Spin selection rule for optical transitions of negatively-charged exciton $X^-$ (see text).} \label{fig1}
\end{figure}

The optical properties of singly charged QDs are dominated by the optical resonances of the negatively-charged exciton $X^-$ (also called trion) which
consists of two electrons bound to one hole \cite{warburton97}. Due to the Pauli's exclusion principle, $X^-$ shows spin-dependent optical
transitions (see Fig. 1b)\cite{hu98}: the left circularly polarized photon (marked by $|L\rangle$ or L-photon) only couples the electron in the
 spin state $|\uparrow\rangle$  to $X^-$ in the spin state $|\uparrow\downarrow\Uparrow\rangle$ with the two electron spins antiparallel; the right
circularly polarized photon (marked by $|R\rangle$ or R-photon) only couples the electron in the spin state $|\downarrow\rangle$ to $X^-$ in the
spin state $|\downarrow\uparrow\Downarrow\rangle$. Here $|\uparrow\rangle$  and $|\downarrow\rangle$ represent electron spin states $|\pm
\frac{1}{2}\rangle$, $|\Uparrow\rangle$ and $|\Downarrow\rangle$ represent heavy-hole spin states $|\pm\frac{3}{2}\rangle$.
The light-hole sub-band and the split-off sub-band are energetically far apart from the heavy-hole sub-band and can be neglected.
The spin is quantized along the normal direction of the cavity, i.e., the propagation direction of the input (or output) light.
This spin selection rule for $X^-$ is also called the Pauli blocking effect \cite{warburton97, calarco03}.

The reflection and transmission coefficients of this $X^-$-cavity structure can be investigated by solving the Heisenberg equations of motion for
the cavity field operator $\hat{a}$ and $X^-$ dipole operator $\sigma_-$, and  the input-output equations\cite{walls94}:
\begin{equation}
\begin{cases}
& \frac{d\hat{a}}{dt}=-\left[i(\omega_c-\omega)+\kappa+\frac{\kappa_s}{2}\right]\hat{a}-\text{g}\sigma_- \\
& ~~~~~~ -\sqrt{\kappa}\hat{a}_{in}-\sqrt{\kappa}\hat{a}^{\prime}_{in} +\hat{H}\\
& \frac{d\sigma_-}{dt}=-\left[i(\omega_{X^-}-\omega)+\frac{\gamma}{2}\right]\sigma_--\text{g}\sigma_z\hat{a}+\hat{G} \\
& \hat{a}_{r}=\hat{a}_{in}+\sqrt{\kappa}\hat{a} \\
& \hat{a}_{t}=\hat{a}^{\prime}_{in}+\sqrt{\kappa}\hat{a} \\
\end{cases}
\label{eq1}
\end{equation}
where $\omega$, $\omega_c$, and $\omega_{X^-}$ are the frequencies of the input photon, cavity mode, and $X^-$ transition, respectively. g
is the $X^-$-cavity coupling strength  given by $\text{g}=(e^2f/4\epsilon_r\epsilon_0m_0V_{eff})^{1/2}$ where $f$ is the $X^-$ oscillator
strength and $V_{eff}$ is the effective modal volume, $\gamma/2$ is the $X^-$ dipole decay rate, and $\kappa$, $\kappa_s/2$ are  the cavity
field decay rate into the input/output modes, and the  leaky modes, respectively. The  background absorption
can also be included in $\kappa_s/2$. $\hat{H}$, $\hat{G}$ are the noise operators related to reservoirs.
$\hat{a}_{in}$, $\hat{a}^{\prime}_{in}$ and $\hat{a}_{r}$, $\hat{a}_{t}$ are the input and output field operators.

In the approximation of weak excitation, i.e., less than one photon inside the cavity per cavity lifetime so that QD is
in the ground state at most time, we take $\langle \sigma_z\rangle \approx -1$. The reflection and transmission
coefficients in the steady state can be obtained
\begin{equation}
\begin{split}
& r(\omega)=1+t(\omega) \\
& t(\omega)=\frac{-\kappa[i(\omega_{X^-}-\omega)+\frac{\gamma}{2}]}{[i(\omega_{X^-}-\omega)+
\frac{\gamma}{2}][i(\omega_c-\omega)+\kappa+\frac{\kappa_s}{2}]+\text{g}^2}.
\end{split}
\label{eq2}
\end{equation}
By taking $\text{g}=0$, we get the reflection and transmission coefficients for an empty cavity
where the QD does not couple to the cavity
\begin{equation}
\begin{split}
& r_0(\omega)=\frac{i(\omega_c-\omega)+\frac{\kappa_s}{2}}{i(\omega_c-\omega)+\kappa+\frac{\kappa_s}{2}} \\
& t_0(\omega)=\frac{-\kappa}{i(\omega_c-\omega)+\kappa+\frac{\kappa_s}{2}}
\end{split}
\label{eq3}
\end{equation}

The reflection and transmission spectra versus the frequency detuning $\omega-\omega_c$ are presented in Fig. 2a for different
coupling strength $\text{g}$. With increasing $\text{g}$ (e.g. by reducing the effective modal volume $V_{\text{eff}}$),
the cavity mode splits into two peaks due to the quantum interference in the \textquotedblleft one dimensional
atom\textquotedblright  regime with $\kappa < 4\text{g}^2/\kappa < \gamma$ \cite{waks06, garnier07} (which has
been experimentally demonstrated recently \cite{englund07}), and the vacuum Rabi splitting in the strong coupling regime
with $\text{g} > (\kappa, \gamma)$ \cite{reithmaier04, yoshie04, peter05, hennessy07, press07, reitzenstein07}.
We notice  that the transmittance or reflectance are different between the empty cavity ($\text{g}=0$) and the coupled
cavity ($\text{g}\neq 0$) (the coupled $X^-$-cavity system is called coupled cavity hereafter). This enables us to make
a photon-spin entangling gate as discussed below.

\begin{figure}[ht]
\centering
\includegraphics* [bb= 91 456 496 752, clip, width=8cm, height=7.0cm]{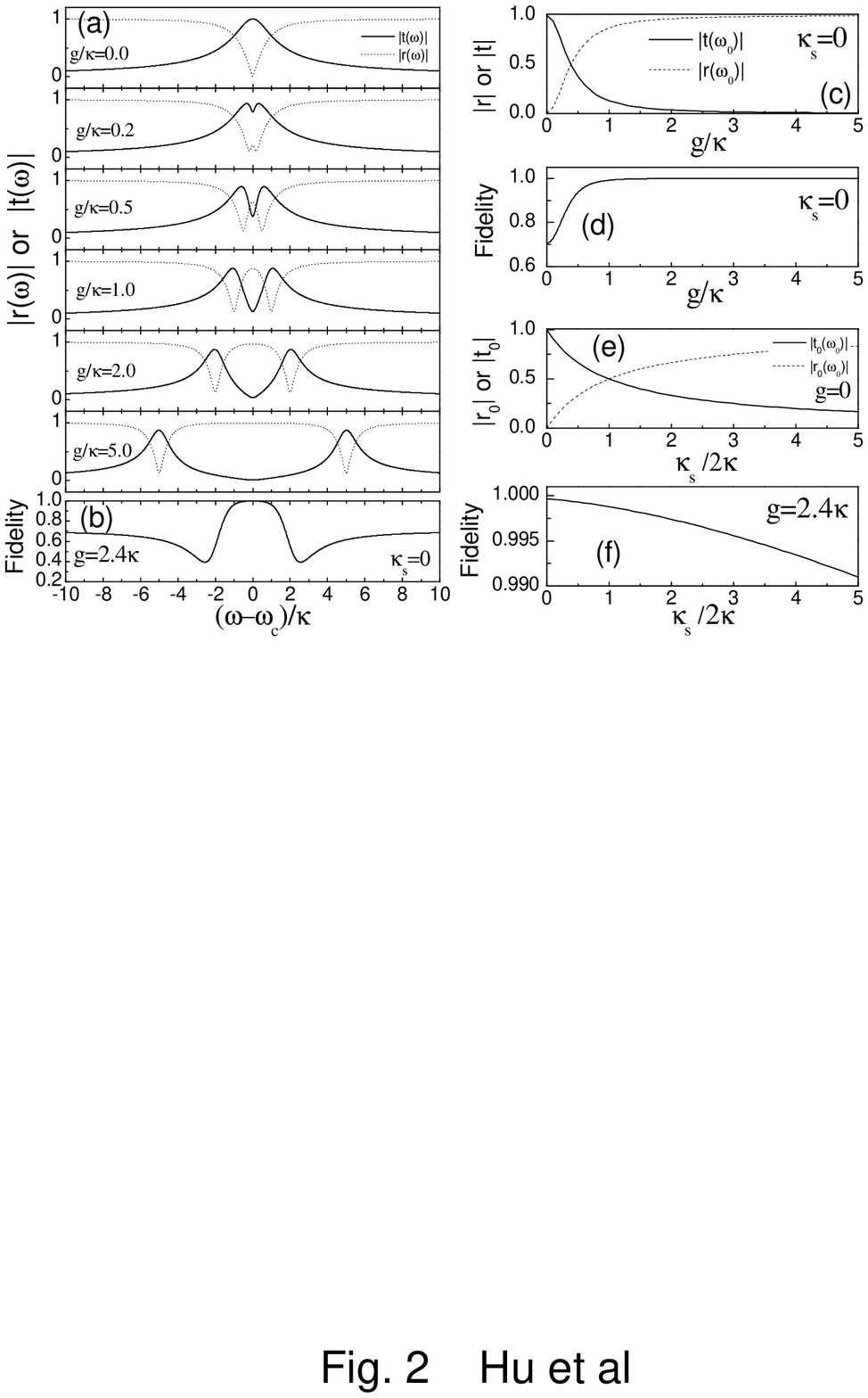}
\caption{Calculated transmission and reflection spectra of the $X^-$-cavity system.
(a) Transmission (solid curves)  and reflection (dotted curves) spectra vs the frequency detuning $(\omega-\omega_c)/\kappa$
for different coupling strength. (b) The gate fidelity vs the frequency detuning in the strong coupling regime
($\text{g}=2.4\kappa$ is taken). High fidelity can be achieved if $|\omega-\omega_c| < \kappa < \text{g}$.
(c) Transmittance $|t(\omega_0)|$ (solid curve)  and  reflectance $|r(\omega_0)|$ (dotted curve) vs the normalized
coupling strength. (d) The  gate fidelity  vs the normalized coupling strength.
(e) Transmittance $|t(\omega_0)|$ (solid curve)  and  reflectance $|r(\omega_0)|$ (dotted curve) vs the normalized side leakage rate.
(f) The  gate fidelity  vs the normalized side leakage rate.
$\omega_c=\omega_{X^-}=\omega_0$ is assumed for (a)-(f). $\kappa_s=0$ and $\gamma=0.1\kappa$ are taken for (a)-(d). } \label{fig2}
\end{figure}

If the single excess electron in the QD lies in the spin state $|\uparrow\rangle$, the L-photon feels a coupled cavity with  reflectance $|r(\omega)|$
and the transmittance $|t(\omega)|$, whereas the R-photon  feels the empty cavity with the reflectance $|r_0(\omega)|$
and transmittance $|t_0(\omega)|$; Conversely, if the electron lies in
the spin state $|\downarrow\rangle$, the R-photon feels a coupled cavity, whereas the
L-photon feels the empty cavity. The difference in transmission and reflection between right and left circularly polarized light,
which can be called giant circular birefringence, means we have created a circular polariser controlled by the electron spin.
For any quantum input we can define a transmission operator
\begin{equation}
\begin{split}
\hat{t}(\omega)=& t_0(\omega)(|R\rangle\langle R|\otimes |\uparrow \rangle\langle \uparrow|+|L\rangle\langle L|\otimes |\downarrow \rangle\langle \downarrow|)\\
& +t(\omega)(|R\rangle\langle R|\otimes |\downarrow \rangle\langle \downarrow|+|L\rangle\langle L|\otimes |\uparrow \rangle\langle \uparrow|),
\end{split}
\label{eq4}
\end{equation}
where $t_0(\omega)$, $t(\omega)$ are the transmission coefficients of the empty cavity and coupled cavity, respectively.

In the strong coupling regime, i.e.,  $\text{g} > (\kappa, \gamma)$ and in the central frequency regime $|\omega-\omega_c|\ll\text{g}$,
we have $|t(\omega)|\rightarrow 0$ (see Fig. 2a), thus the transmission operator
can be simplified as
\begin{equation}
\hat{t}(\omega)\simeq t_0(\omega)(|R\rangle\langle R|\otimes |\uparrow \rangle\langle \uparrow|+|L\rangle\langle L|\otimes |\downarrow \rangle\langle \downarrow|).
\label{eq5}
\end{equation}
Obviously, this transmission operator is now constructed from the empty cavity only.
We show later how this operator can be used as a photon-spin entangling gate. Here however we can define an operator
fidelity (based on amplitude) from equations (\ref{eq4}) and (\ref{eq5}) as
\begin{equation}
F=\frac{|t_0(\omega)|}{\sqrt{|t_0(\omega)|^2+|t(\omega)|^2}}.
\label{fid1}
\end{equation}
Near-unity fidelity is reached when $|t(\omega)|\rightarrow 0$ which is only achieved within a small frequency window $|\omega-\omega_c|<\kappa$
(see Fig. 2b) and in the strong coupling regime with $\text{g}>(\kappa, \gamma)$ (see Fig. 2c and Fig. 2d).
The strongly coupled QD-cavity has been demonstrated in various microcavities and nanocavities
recently  \cite{reithmaier04, yoshie04, peter05, hennessy07, press07, reitzenstein07}.
For micropillars with diameter around $1.5~\mu$m, the coupling strength $\text{g}=80~\mu$eV
and the quality factor more than $4\times 10^4$ (corresponding to
$\kappa=33~\mu$eV) have been reported \cite{reithmaier04, reitzenstein07}, indicating $\text{g}/\kappa=2.4$  is achievable
for the In(Ga)As QD-cavity system. $\gamma$ is about several $\mu$eV.  Our calculations in Fig. 2 are based on
these experimental parameters.

A practical optical cavity can have some side leakage, which induces a decrease in the transmittance of the empty cavity  and the gate fidelity
(see Fig. 3e and Fig. 3f).  However, the improvement of fabrication techniques can  suppress the side leakage \cite{reitzenstein07}.
When the side leakage is made negligible compared with the main cavity decay into the input/output modes, we get $|t_0(\omega_0)|=1$ and
unity gate fidelity.

For a realistic QD, the spin selection rule discussed earlier is not perfect if we take the
heavy-light hole mixing into account.  This can reduce the gate fidelity by a few percent as the hole mixing in the valence band
is in the order of a few percent \cite{bester03, calarco03} [e.g., for self-assembled In(Ga)As QDs]. The hole mixing  could  be
reduced by engineering the shape and size of QDs or using  different types of QDs.

As discussed above, the photon-spin entangling gate requires the weak excitation condition, i.e., the input light intensity has to be
less than one photon per cavity lifetime. This condition can be satisfied by single photons, e.g. QD single photon sources which
can be triggered electrically or optically \cite{moreau01, yuan02, santori02}.
Recently there are lots of experimental efforts to develop high-quality QD single-photon sources with high efficiencies,
small multi-photon events and time-bandwidth limited photon pulses \cite{shields07}.

This photon-spin entangling gate can also work in  the reflection geometry, but its application is more complicated and we leave the discussions
elsewhere. As a result, the photon-spin entangling gate in the transmission geometry  is only $50\%$ efficient.

In the following,  we show that this  photon-spin entangling gate  can be used for QND measurement of a single electron spin, and also can work as
photon-spin entangler, spin entangler, or  photon entangler. With this gate,  reversible quantum state transfer between photon and spin can be
implemented. Compared with our previous gate  \cite{hu08a, hu08b} and Turchette et al's  conditional
phase shift gate using a single-sided cavity \cite{turchette95}, this photon-spin entangling gate using a double-sided cavity
is more robust and flexible. We notice  that other photon-spin entangling gates was also reported recently \cite{flindt07, hu08a, hu08b, lindner08}.

\section{Single-shot optical  QND measurement of a single spin}

If we prepare the input
photon in a linear polarization state $|H\rangle=(|R\rangle+|L\rangle)/\sqrt{2}$ and the electron spin  in the
state $|\psi^s\rangle=|\uparrow\rangle$, according to equation (\ref{eq5}) the state transformation is
\begin{equation}
|H\rangle \otimes |\uparrow\rangle \xrightarrow{\hat{t}(\omega)} \frac{t_0(\omega)}{\sqrt{2}}|R\rangle |\uparrow\rangle.
\end{equation}
So only the right-handed circularly polarized component is transmitted (see Fig. 3a).  Similarly, if the electron spin is in the
state $|\downarrow\rangle$, only the left-handed circularly polarized component is transmitted (see Fig. 3b). Obviously,
this is a circular polariser which allows only one circular polarized light to be
transmitted depending on the spin state. This feature enables us to detect the electron spin by measuring the helicity of the transmitted
light using a $\lambda/4$ wave plate and a polarizing beam splitter (see Fig. 3c).

\begin{figure}[ht]
\centering
\includegraphics* [bb= 137 249 443 553, clip, width=6cm, height=6cm]{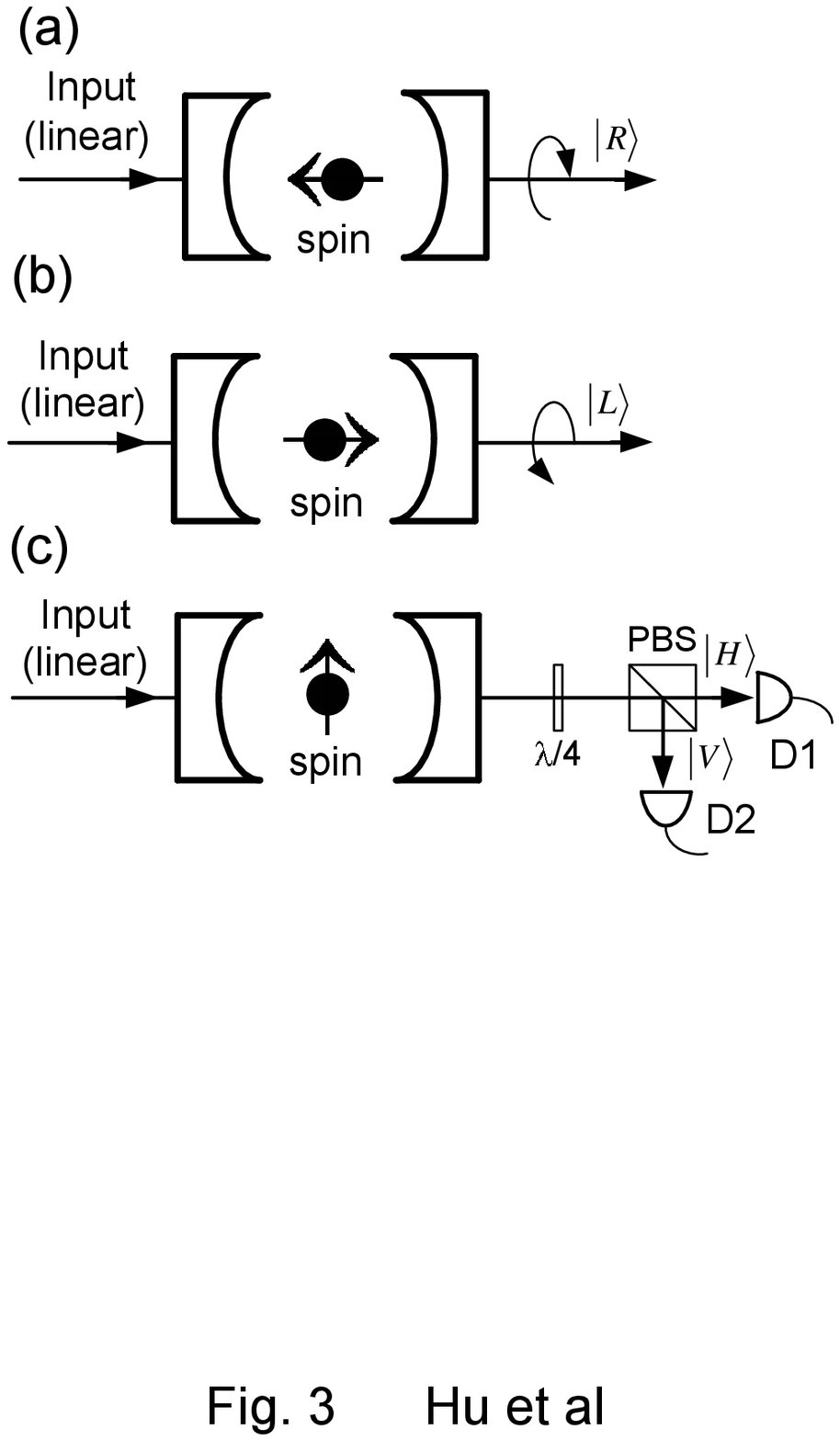}
\caption{QND measurement of a single QD spin. (a) The right-circularly polarized component of a linearly polarized light
is transmitted if the electron spin in the  $|\uparrow\rangle$ state. (b) The left-circularly polarized component of a linearly polarized
light is transmitted if the electron spin in the  $|\downarrow\rangle$ state. (c) Both the right- and left-circularly polarized component of a
linearly polarized light are transmitted if the electron spin in a superposition state. PBS (polarizing beam splitter), D1 and D2 (photon
detectors), and $\lambda/4$ (quarter-wave plate).  } \label{fig3}
\end{figure}

If the electron spin is in an arbitrary superposition state $|\psi^s\rangle=\alpha |\uparrow\rangle+\beta|\downarrow\rangle$ (see Fig. 3c),
the state transformation is
\begin{equation}
|H\rangle \otimes (\alpha |\uparrow\rangle+\beta|\downarrow\rangle) \xrightarrow{\hat{t}(\omega)} \frac{t_0(\omega)}{\sqrt{2}}
(\alpha|R\rangle |\uparrow\rangle+\beta|L\rangle |\downarrow\rangle).
\end{equation}
Thus after transmission, the light polarization state becomes entangled with the spin state. This is why we call this gate a photon-spin
entangling gate.  If we measure the light in $|R\rangle$ (or $|L\rangle$) polarization, the electron spin
collapses to  $|\uparrow\rangle$ (or $|\downarrow\rangle$) state.  Although this gate work in the near resonance region, the weak excitation condition
means nearly no real excitation occurs in the  $X^-$-cavity system. As a result, the disturbance  to the electron spin system due to the light input
is quite small.
Within the spin relaxation time ($\sim $ms) \cite{kroutvar04}, repeated measurements will yield the same results, so this single-shot spin
detection  method is a QND measurement \cite{grangier98}, in contrast to other single-spin detection methods by the time-averaged Faraday rotation or Kerr rotation measurement reported recently \cite{berezovsky06, atature07}. In parallel, a QND  measurement of  single photon polarization state
could also be implemented using the above spin QND measurement. QND  measurement is critical for scalable quantum information
processing \cite{liu05, nemoto05}.

The QD spin eigen-state can be prepared, for example, by optical spin pumping \cite{atature06, xu07}. From the above discussions,
we see the single-shot QND measurement of single spin can be also used to prepare the spin eigen state and cool the spin via photon detection \cite{liu05}.
From the spin basis state, there are two ways to create the spin superposition state: either via spin-flip Raman transitions \cite{atature06}, or by
performing single spin rotations using nanosecond ESR microwave pulses \cite{petta05}. Recently,  ultrafast optical coherent control of
electron spins has been reported in quantum wells on femtosecond time scales \cite{gupta01} and in  QDs on
picosecond time scales \cite{berezovsky08, press08}, which is much shorter than the QD spin coherence time ($T_2\sim \mu$s).
This allows ultrafast $\pi/2$ spin rotation which is required in our schemes for spin state
preparation or spin Hadamard operation.

\section{Entangle remote spins via a single photon}

 We show here that the photon-spin entangling gate can be used to generate entanglement between remote spins in different
cavities via a single photon (see  Fig. 4a). In the first $X^-$-cavity system, the  spin is prepared in the  state
$|\psi^s\rangle_1=\alpha_1 |\uparrow\rangle_1+ \beta_1|\downarrow\rangle_1$ and transmission operator is  $\hat{t}_1(\omega)$; In
the second $X^-$-cavity system, the  spin  is prepared in the  state $|\psi^s\rangle_2=\alpha_2 |\uparrow\rangle_2+
\beta_2|\downarrow\rangle_2$ and transmission operator is  $\hat{t}_2(\omega)$. Both $X^-$-cavity systems work in the strong coupling regime
to get high gate fidelity, but the parameters $\text{g}$, $\kappa$, $\kappa_s$, $\omega_c$ and $\omega_{X^-}$ for this two
systems are not necessary to be the same.

A single photon in $|H\rangle$ polarization passes through the first cavity, then through the second cavity, after which its polarization is checked (see Fig. 4a).  The corresponding state transformation is
\begin{equation}
\begin{split}
& |H\rangle\otimes  (\alpha_1 |\uparrow\rangle_1+ \beta_1 |\downarrow\rangle_1)\otimes (\alpha_2 |\uparrow\rangle_2+ \beta_2 |\downarrow\rangle_2)\xrightarrow{\hat{t}_{1,2}(\omega)}\\
&  \frac{t_{10}(\omega)t_{20}(\omega)}{\sqrt{2}}
(\alpha_1\alpha_2|R\rangle|\uparrow\rangle_1|\uparrow\rangle_1+\beta_1\beta_2|L\rangle|\downarrow\rangle_1|\downarrow\rangle_2)
\end{split}
\label{se1}
\end{equation}
By applying the Hadamard gate on the photon state using a polarizing beam splitter, we obtain entangled spin states
\begin{equation}
|\Phi^s_{12}\rangle=\alpha_1\alpha_2|\uparrow\rangle_1|\uparrow\rangle_2\pm\beta_1\beta_2|\downarrow\rangle_1|\downarrow\rangle_2
\end{equation}
on detecting the photon in the $|H\rangle$ state (for \textquotedblleft +\textquotedblright), or in  $|V\rangle=(|R\rangle-|L\rangle)/\sqrt{2}$
state (for \textquotedblleft -\textquotedblright).
On setting the coefficients $\alpha_{1,2}$ and $\beta_{1,2}$ to $1/\sqrt{2}$, we get maximally entangled spin states.

We see the single photon works as a quantum bus to couple or entangle remote spins on demand, but the two spins
in two cavities can be slightly different in their transition frequencies.
However, if the cavity mode frequency $\omega_c$ and the $X^-$ transition frequency $\omega_{X^-}$ match  with the photon frequency $\omega$
for the two $X^-$-cavity systems, the success probability $|t_{10}(\omega)t_{20}(\omega)|^2/2$ to achieve the spin entanglement
can be increased. As discussed earlier, if the side leakage
can be made significantly small, $|t_{10}(\omega)|$ and $|t_{20}(\omega)|$ can both reach unity and we get the maximal success probability of $50\%$.
But we know for certain we have succeeded in entangling the spins when a photon is detected.
The schemes based on quantum interference of emitted photons can generate remote atomic entanglement \cite{chou05, moehring07}, and could be
extended to entangle distant spins \cite{childress06, simon07a}.  However these schemes suffer from low success probability, and require
identical atoms or spins \cite{moehring07}. There are also some other  schemes based on Faraday rotation \cite{leuenberger05, hu08a} and the probabilistic schemes based on the dispersive spin-photon interactions \cite{grond08} using bright coherent light as proposed by van Loock et al and Ladd et al \cite{loock06}.

The above scheme can be easily extended to generate multi-spin entangled states, such as Greenberger-Horne-Zeilinger (GHZ) states \cite{greenberger90} by passing
the single photon through all cavities and finally checking the photon polarization. On setting all $\alpha's$ and $\beta's$ to $1/\sqrt{2}$,
we get maximally entangled spin GHZ states:
\begin{equation}
|\text{GHZ}^s\rangle_N=\frac{1}{\sqrt{N}}(|\uparrow\rangle_1|\uparrow\rangle_2\cdot\cdot\cdot |\uparrow\rangle_N \pm |\downarrow\rangle_1|\downarrow\rangle_2\cdot\cdot\cdot |\downarrow\rangle_N)
\end{equation}
Alternatively, starting  from  entangled spin pairs, we could build higher-order entangled spin states such as GHZ states or
cluster states \cite{briegel01} with N unlimited. The success probability is $1/2^k$ depending on the number k of single photons
used. Again the detection of the photons heralds a successful entanglement operation.

We point out here that the influence of photon reflection between cavities can be removed by utilizing suitable timing system.
Once we have created entangled spin states, either optical or electrical pumping can be used to excite $X^-$ in  QDs.
Spin entanglement  is then  transferred to photon polarization entanglement via $X^-$ emissions  due to the same optical spin
selection rule of $X^-$ as discussed earlier. However, we show another scheme below -
a photon entangler which can entangle independent photons with different frequencies or different pulse length.

\begin{figure}[ht]
\centering
\includegraphics* [bb= 65 359 536 589, clip, width=8cm, height=4.5cm]{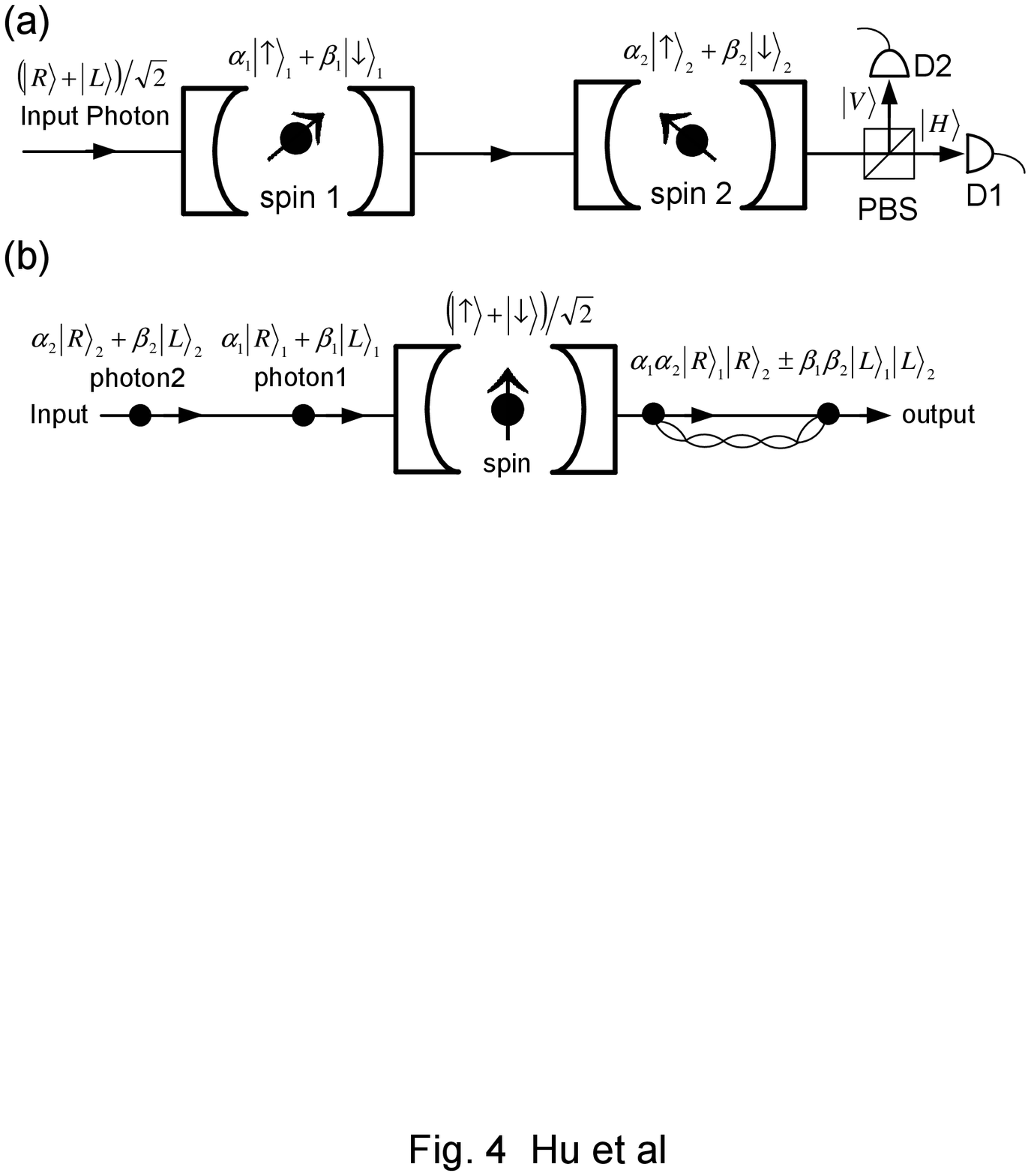}
\caption{Schematic diagram of a spin / photon entangler. (a) A proposed scheme to entangle remote spins in different
microcavities via a single photon. PBS (polarizing beam splitter) and D1 and D2 (photon detectors).
(b) A proposed scheme to entangle independent photons via a single  spin in a microcavity. } \label{fig4}
\end{figure}

\section{Entangle independent photons via a single spin}

As shown in Fig. 4b, photon 1 in the state $|\psi^{ph}\rangle_1=\alpha_1|R\rangle_1+\beta_1|L\rangle_1$ and photon 2 in the state
$|\psi^{ph}\rangle_2=\alpha_2|R\rangle_2+\beta_2|L\rangle_2$ are input into the cavity in sequence. The two independent photons can
have different frequencies, but both are in the frequency window $|\omega-\omega_c|<\kappa$. The electron spin is prepared in a superposition
state $|\psi^{s}\rangle=\frac{1}{\sqrt{2}}(|\uparrow\rangle+|\downarrow\rangle)$.
The transmission operator $\hat{t}(\omega)$ for the $X^-$-cavity
system is again described by equation (\ref{eq5}).

After transmission, the state transformation is
\begin{equation}
\begin{split}
(\alpha_1 & |R\rangle_1  +\beta_1|L\rangle_1) \otimes (\alpha_2|R\rangle_2+\beta_2|L\rangle_2) \otimes |\psi^{s}\rangle \\
~ \xrightarrow{\hat{t}(\omega)} & \frac{t_0(\omega_1)t_0(\omega_2)}{\sqrt{2}} \left(\alpha_1\alpha_2|R\rangle_1|R\rangle_2|\uparrow\rangle+\beta_1\beta_2|L\rangle_1|L\rangle_2|\downarrow\rangle \right).
\end{split}
\label{pe1}
\end{equation}

By applying a Hadamard gate on the electron spin (e.g., using a $\pi/2$ microwave or optical pulse), the right side of equation (\ref{pe1}) becomes
\begin{equation}
\begin{split}
 \frac{t_0(\omega_1)t_0(\omega_2)}{2}&\{(\alpha_1\alpha_2|R\rangle_1|R\rangle_2+\beta_1\beta_2|L\rangle_1|L\rangle_2)|\uparrow\rangle \\
& +(\alpha_1\alpha_2|R\rangle_1|R\rangle_2-\beta_1\beta_2|L\rangle_1|L\rangle_2)|\downarrow\rangle\}
\end{split}
\label{pe2}
\end{equation}

Next, the electron spin eigen-state can be detected by the QND measurement as discussed earlier using a weak coherent light (or
single photons) in H polarization. Depending on the detected spin state in $|\uparrow\rangle$ or $|\downarrow\rangle$, we get the
entangled photon states
\begin{equation}
\Phi^{ph}_{12}=(\alpha_1\alpha_2|R\rangle_1|R\rangle_2 \pm \beta_1\beta_2|L\rangle_1|L\rangle_2)
\label{pe3}
\end{equation}
On setting the coefficients $\alpha_{1,2}$ and $\beta_{1,2}$ to $1/\sqrt{2}$, maximally entangled photon states can be generated.

Although photon 1 and photon 2 never meet before, each of them gets entangled with the electron spin after sequentially interacting
with the spin. The spin measurement then projects the two photons into entangled states.  This  entanglement-by-projection
scheme does not require photon indistinguishability or
photon interference as demanded by other schemes using photon mixing on a beam splitter \cite{fattal04}.
This kind of  single-photon pulses  can come from  QD single photon
sources \cite{moreau01, yuan02, santori02}.

Recent experiments have shown GaAs or
In(Ga)As single QDs have long electron spin coherence time ($T_2\sim \mu$s)\cite{petta05, greilich06} and  spin relaxation
time ($T_1\sim$ms) \cite{kroutvar04}. Due to the spin decoherence,   the density matrix of the electron spin in the initial
state $|\psi^{s}\rangle=\frac{1}{\sqrt{2}}(|\uparrow\rangle+|\downarrow\rangle)$ evolves at time t ($t\ll T_1$)
\begin{equation}
\rho(t)=\begin{pmatrix}
1/2 & e^{-t/T_2}/2  \\
e^{-t/T_2}/2 &  1/2
\end{pmatrix},
\end{equation}
which represents a spin mixed state. As a result, the entanglement fidelity with respect to equation (\ref{pe3})  becomes
\begin{equation}
F=\frac{1}{2}(1+e^{-t/T_2}),
\end{equation}
which decreases with t. Therefore high fidelity photon entanglement can only be achieved when the time
interval  between two photons is much shorter than the spin coherence time ($T_2\sim \mu$s) in the QD.
This entanglement between photons with different arrival time is ideal for quantum relay type applications.

If increasing $|t_0(\omega)|$ to one  by optimizing the cavity, the success probability for the photon entanglement generation
can reach $25\%$, so coincidence measurement of photons is required to post-select the entangled state.

We could also extend this scheme to generate multi-photon GHZ states by passing all photons through the cavity in sequence
and finally checking the spin state after applying a Hadamard gate on the spin. An alternative way to generate GHZ \cite{greenberger90} or
cluster states \cite{briegel01} is to start from the generation of entangled photon pairs and then repeat this procedure
to increase the size such that the photon number N can be unlimited.
On setting all $\alpha's$ and $\beta's$ to $1/\sqrt{2}$, we get maximally entangled photon GHZ states:
\begin{equation}
|\text{GHZ}^{ph}\rangle_N=\frac{1}{\sqrt{N}}(|R\rangle_1|R\rangle_2\cdot\cdot\cdot |R\rangle_N \pm |L\rangle_1|L\rangle_2\cdot\cdot\cdot |L\rangle_N).
\end{equation}
The maximal success probability is then $1/2^N$.

\begin{figure}[ht]
\centering
\includegraphics* [bb= 119 440 449 676, clip, width=7cm, height=5cm]{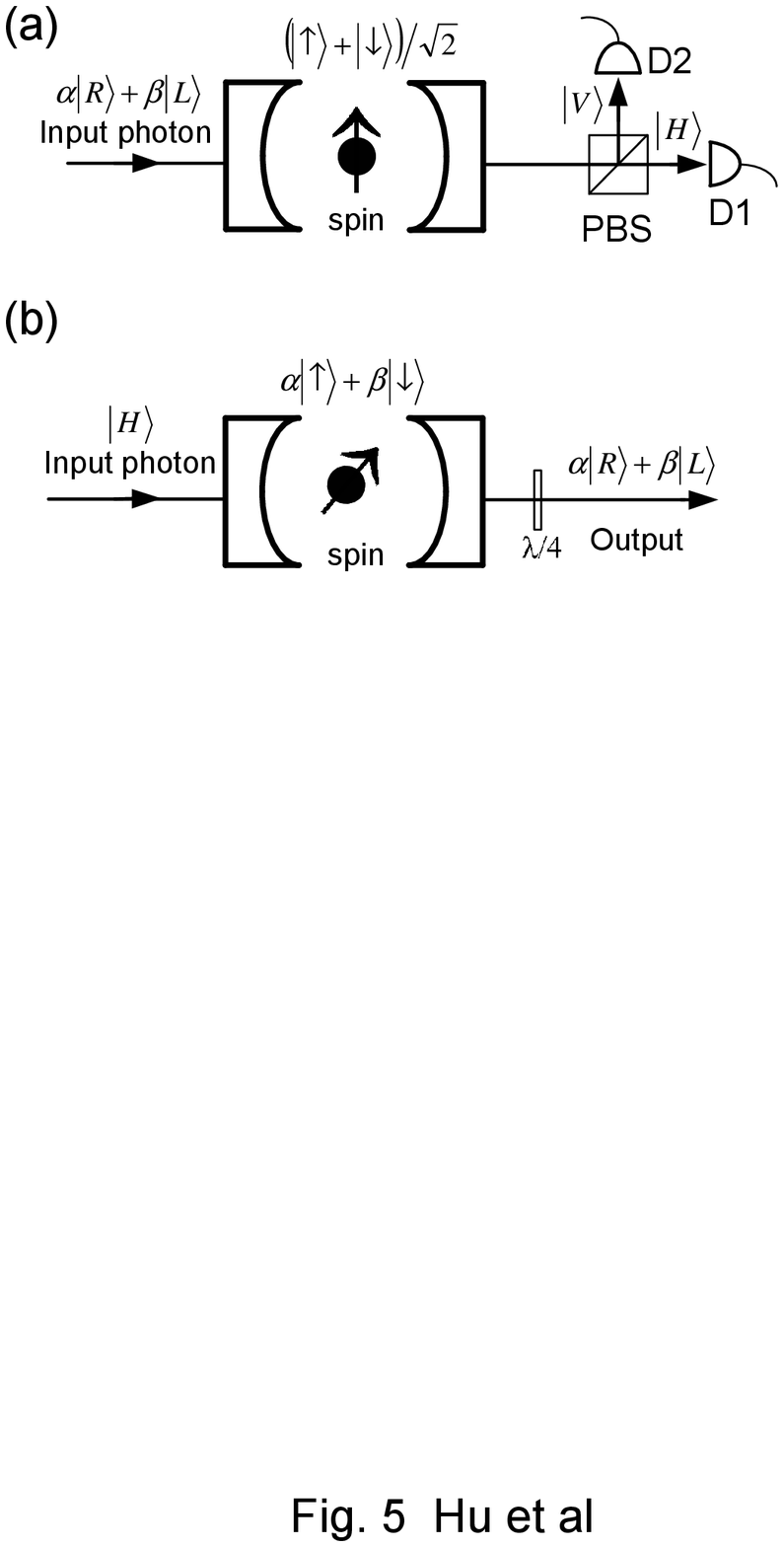}
\caption{Schematic diagram of a photon-spin quantum interface. (a) State transfer from a photon to a spin. (b) State transfer
from a spin to a photon. PBS (polarizing beam splitter), D1 and D2 (photon detectors), and $\lambda/4$ (quarter-wave plate).} \label{fig5}
\end{figure}

\section{Photon-spin quantum interface}

Quantum interface is a critical component for quantum networks. Here we show reversible and coherent quantum-state transfer
between photon and spin using the photon-spin entangling gate.
In Fig. 5a, a photon in an arbitrary state $|\psi^{ph}\rangle=\alpha|R\rangle+\beta|L\rangle$ is input to the cavity with
the electron spin prepared in the state $|\psi^{s}\rangle=\frac{1}{\sqrt{2}}(|\uparrow\rangle + |\downarrow\rangle)$.
After transmission, the photon and the spin  become entangled, i.e,
\begin{equation}
(\alpha|R\rangle+\beta|L\rangle) \otimes |\psi^{s}\rangle
\xrightarrow{\hat{t}(\omega)} \frac{t_0(\omega)}{\sqrt{2}} \left(\alpha |R\rangle |\uparrow\rangle+\beta |L\rangle|\downarrow\rangle\right).
\label{ps1}
\end{equation}
By applying a Hadamard gate on the photon state  using a polarizing beam splitter, we obtain a spin state
$|\Phi^s_1\rangle=\alpha|\uparrow\rangle\pm\beta|\downarrow\rangle$ on detecting a photon in the
$|H\rangle$ or $|V\rangle$ state. Therefore, the photon state is transferred to the electron spin state.

In Fig. 5b, a photon in the polarization state
$|\psi^{ph}\rangle=(|R\rangle+|L\rangle)/\sqrt{2}$ is input to the cavity with the electron spin in an arbitrary state
$|\psi^{s}\rangle=\alpha |\uparrow\rangle+\beta |\downarrow\rangle$.
After transmission, the photon  and the spin  become entangled, i.e,
\begin{equation}
|\psi^{ph}\rangle \otimes (\alpha |\uparrow\rangle+ \beta |\downarrow\rangle) \xrightarrow{\hat{t}(\omega)}
\frac{t_0(\omega)}{\sqrt{2}}\left(\alpha|R\rangle|\uparrow\rangle+\beta |L\rangle|\downarrow\rangle \right).
\label{trans7}
\end{equation}
After  applying a Hadamard gate on the electron spin (e.g., using a $\pi/2$ microwave or optical pulse), the spin eigen-state is detected by the
QND measurement as discussed earlier. On detecting the electron spin in the  $|\uparrow\rangle$ or $|\downarrow\rangle$ state, the photon
is then projected in the state $|\Phi^{ph}_1\rangle=\alpha|R\rangle\pm\beta|L\rangle$.  So the spin state is transferred
to the photon state.

In contrast to the original teleportation protocol which involves three qubits \cite{bennett93}, our state transfer scheme requires only two
qubits thanks to the tunable amount of  entanglement. The success probability is $|t_0(\omega)|^2/2$, which can be increased to $50\%$ by
optimizing the cavity. The state transfer fidelity is determined by the gate  fidelity as described by equation (\ref{fid1}).

\section{Conclusions}

Entanglement is a fundamental resource in quantum information science. With the proposed photon-spin entangling gate, it
is possible to generate almost all kinds of local or remote entanglement among photons and QD spins with high fidelity.  This entanglement
would find wide applications in quantum communications such as  quantum cryptography and quantum teleportation. Moreover, this entanglement
is essential to implement a quantum bus, quantum interface, quantum memories and  quantum repeaters, all of which are critical building blocks
for quantum networks. The high-order multiparticle entanglement could be used for entanglement-enhanced quantum measurement \cite{giovannetti04},
or cluster-state based quantum computing \cite{raussendorf01, nielsen06}.

This gate can also work as an active device such as a polarization-controlled single photon source \cite{moreau01, yuan02, santori02}, which could be driven by the electron spin dynamics. These single photons on demand can be sent back to the gate to get entangled photons based on our schemes.
Techniques for manipulating single photons have been well developed, and significant progress on fast QD-spin cooling
and manipulating has been made recently \cite{atature06, xu07, petta05, berezovsky08}. Together with this work, we believe
a charged QD in an optical cavity is promising for solid-state quantum networks and quantum computing.

\section*{Acknowledgements}

C.Y.H. thanks M. Atat\"{u}re, S. Bose, and S. Popescu for helpful discussions.
J.G.R. acknowledges support from the Royal Society.  This work is partly funded by EPSRC-GB IRC in Quantum
Information Processing, QAP (Contract No. EU IST015848), and MEXT from Japan.

\end{document}